\documentclass[%
 reprint,
superscriptaddress,
 amsmath,amssymb,
 aps,
]{revtex4-2}

\usepackage{graphicx}% Include figure files
\usepackage{dcolumn}% Align table columns on decimal point
\usepackage{bm}% bold math
\usepackage{braket}
\usepackage{comment}
\usepackage{amsmath}
\usepackage{amssymb}   % for math
\usepackage{mciteplus} % for advanced citations
\mciteSetMidEndSepPunct{\space}{\ifmciteBstWouldAddEndPunct.\else\fi}{\relax} %separates citations with a space.
\usepackage{lmodern} % for German related special characters
\usepackage{braket}
\usepackage{siunitx}
\begin{document}

\preprint{APS/123-QED}

% the following line is for submission, including submission to the arXiv!!
%\hspace{5.2in} \mbox{Fermilab-Pub-04/xxx-E}

\title{Coherent Field Emission Upon Ultrafast Laser Irradiation of the Tip Plasmon}
\author{Joonhee~Lee}\affiliation{Department of Physics, University of Nevada, Reno, Nevada, 89557, USA}
\author{Shawn M.~Perdue}\affiliation{Department of Chemistry, University of California, Irvine, California 92697-2025, USA}
\author{Alejandro~Rodriguez Perez}\affiliation{Department of Chemistry, University of California, Irvine, California 92697-2025, USA}
\author{V. A.~Apkarian}\affiliation{Department of Chemistry, University of California, Irvine, California 92697-2025, USA}

\date{\today}

\begin{abstract}
Irradiation of sharp silver tips with femtosecond laser pulses leads to photoassisted coherent field emission without a static field. We reconstruct the time profile of the emission, and show that the process is entirely governed by the collective response of the tip plasmon and its field emission. Weak-field optical excitation leads to multiphoton absorption and field emission from the tip apex due to the enhanced local field. The attendant sharp field gradient ensures ponderomotive acceleration of emitted electrons and non-local light-matter interaction. The crossover regime in which simultaneous multiphoton absorption and optical field emission take place is evidenced by the time profile of electron emission correlation, laser power dependence, and polarization angle dependence of each harmonic current.
\end{abstract}

\maketitle

Investigations into the governing physics of ultrafast laser driven electron emission from sharp metal tips \cite{Hommelhoff2006, Barwick, Ropers2007, Ropers20072, Hilbert, Yanagisawa2010, Bormann, Wu, Matte} are motivated by applications that seek ultimate limits in joint space-time resolution, as in time-resolved implementations of electron microscopy \cite{Yang}, electron diffraction \cite{Aidelsburger}, and scanning tunneling microscopy (STM) \cite{Lee, Garg}. At typical photon energies $\phi/\hbar\omega\geq3$ (where $\phi$ is the workfunction of the metal) and where radiation couples to the metal plasmon \cite{Petek}, field emission with spatial localization on atomic scales has been demonstrated \cite{Hommelhoff2006, Yanagisawa2010}. Sub-femtosecond temporal response has been established by demonstrating above threshold photoemission \cite{Hommelhoff2011} and barrier tunneling \cite{Garg} sensitive to the carrier envelope phase of the laser pulse. This strong-field effect had been predicted to be operative on plasmonic structures \cite{Stockman}. However, the experimental measurements reveal a variety of operative mechanisms. In particular, when using large static extractor fields, tunneling emission from multiphoton excited electrons is observed to dominate \cite{Yanagisawa2011}. Also, a framework explaining the transition from multiphoton absorption to strong-field tunneling in terms of photon numbers was reported \cite{Bormann}. To reach ultimate limits in spatio-temporal localization, phase coherent field emission is required. Here, we describe interferometric measurements of photo-induced field emission that retains the optical phase of the incident light, dominated by tunneling of the quasiparticle tip plasmon.

Our measurements are carried out on silver tips, which find their utility in tip-enhanced Raman spectroscopy (TERS) \cite{Leenature} due to sustaining bright plasmons in the optical spectral range. Through either propagating modes \cite{Raschke2010} that undergo superfocusing \cite{Nerkararyan2000} or direct excitation of the apex mode, tip plasmons \cite{Anderson2010} provide the primary channel of light-matter coupling. The coherent collective exciation of surface electrons (plasmon) arises as the screening response to an applied optical field, which is most naturally understood as the surface waves on a jellium by shaking the bucket that holds it, that is modulated by a time-harmonic external potential, $\delta \rho(r;\omega_o) = \int \chi(r,r';\omega_o)V_{ext}(r';\omega_o)dr'$, where $\chi$ is the density-density susceptibility and the time-harmonic $V_{ext}(w_0)$ acts to bend the image potential \cite{Kempa}. At frequencies much lower than the surface plasmon resonance (4.2 eV for Ag), $\delta \rho$ approximates the static limit obtained in the classic work of Lang and Kohn \cite{Lang}: The centroid of the screening charge density peaks in the spill-out region of the jellium and decays into the bulk along the surface normal by Friedel oscillation, $\delta \rho(z)=\delta \rho_0 \sin(2k_F z)/2k_F z$, as a quasiparticle of momentum $k_{Fr}=\sqrt{2mE_F}/\hbar$. Sufficiently large fields bend the vacuum level to generate a Schottky barrier, through which the quasiparticle plasmon may tunnel, to field emit electrons \cite{Ohwaki}. In an optical field, the coherent field emission follows the optical phase, which we resolve in this report. We resolve the time profile of the emission and its angular distribution through the nonlinear implementation of the cross-polarized double-beat (CPDB) method \cite{Lee} (See supplement). 

The power dependence, time profile, energy distribution, and polarization dependent angular distribution of the emission identify the dominant channel to be photo-assisted tunneling and subsequent ponderomotive accereration of the harmonics of the coherence prepared by the field-modulated image potential. The oscillating potential changes the energy $E$ of electrons in the jellium into a set of $E+n\hbar\omega_o$ at harmonics of the driving frequency, $n=0, \pm1, \pm2, \cdots$, described by Tien and Gordon's Floquet states, $\rho(E)=\sum_n \rho_0(E+n \hbar\omega_o)J_n(eV_{ext}/\hbar \omega_0)$, which consist of copies of the unperturbed electron density $\rho_0$ as harmonic sidebands with rapidly dropping amplitude as a function of order $n$, given by the Bessel function of the first kind, $J_n(eV_{ext}/\hbar \omega_0)$ \cite{Tien, ButtikerPRL, Buttiker}. Each harmonic acts as a quasiparticle of Friedel momentum $\sqrt{2m(E+n\hbar\omega_o)}/\hbar$, subject to successively diminishing tunneling barriers. Staircase excitation can be directly decomposed in the interferometric measurements and the diminishing barriers with harmonic order is evident in the envelop of the emitted wavepackets and in the angular distribution of photocurrent in cross-polarization, as we expand below.

The time profile establishes that linear excitation leads to nonlinear emission that tracks the optical phase. The ejected electrons are directional and their final energy distribution is consistent with ponderomotive acceleration \cite{Dombi}, obviating the need for extractor fields. Their angular distribution establishes that the nonlinearity arises from the large amplitude displacement of plasmons over the steep gradient of the local field. The emission is consistent with the mechanism of "ponderomotive ionization" anticipated through prior theory \cite{Faisal2005}.

The measurements are carried out in an ultrahigh vacuum STM on silver tips annealed \textit{in situ}. The tip is retracted well outside (gap $>$ 1 $\mu$m) the typical STM gap distance (a few $\si{\angstrom}$) to ensure that the detected current is strictly due to field emission. Unless stated otherwise, the current is measured with a +2 V sample bias with respect to the tip to ensure emitted electrons do not experience an electrostatic barrier towards the substrate. The substrate is a NiAl(110) single crystal. The detailed optical setup is described in the supplementary material. 

\begin{figure}[b]
\centering
\includegraphics[width=1\columnwidth]{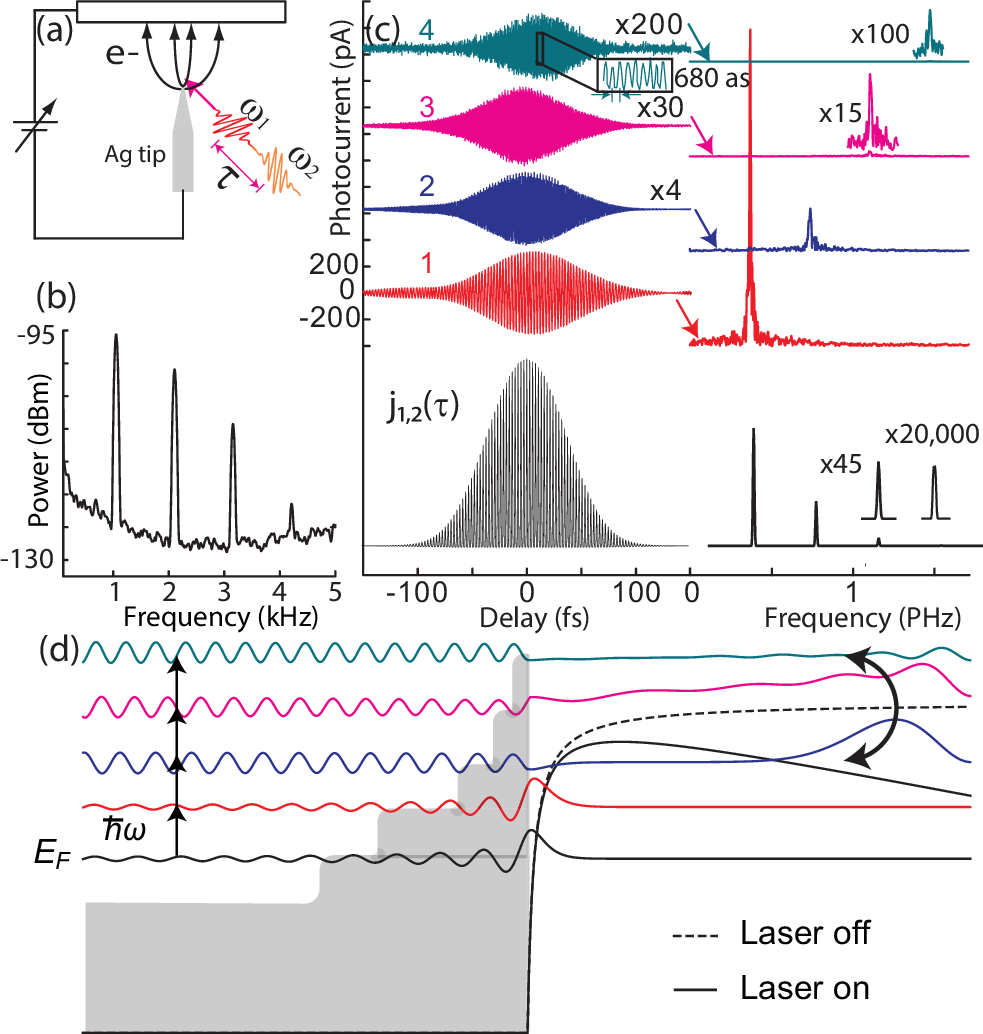}
\caption{\label{fig:1} (a) Schematic of the experiment: Electron emission measured under irradiation of the STM junction by cross-polarized pulse trains. (b) Typical spectrum of the current shows a progression of four harmonics of the double-beat frequency ($\sim$1 kHz). (c) Left panel: Cross-correlation current as a function of delay between pulse trains, recorded at harmonics of the double-beat. Right panel: Fourier transform of the time correlations showing harmonics of the optical frequency ($\lambda$ = 800 nm). The reconstruction of the emission time-profile (in black) is obtained by forward simulation using Eq. 1. (d) Snapshot of charge density difference of harmonics.}
\end{figure}

Briefly, two cross-polarized pulse trains obtained from a Ti$^+$:Sapphire laser ($\lambda=800$ nm, pulsewidth $\sim$50 fs before entering the UHV chamber) are recombined in a Mach-Zehnder interferometer, after frequency shifting one of the arms with an acousto-optic modulator (AOM). The collinear beams are focused on the STM-junction with a plano-convex lens (focal length = 10 cm, nominal spot size = 25 $\mu$m) at $45^{\circ}$ incidence relative to the tip z-axis (Fig. 1(a)). The cross-polarized beams mix through the anisotropic susceptibility of the tip plasmon to create a linear interference (electronic coherence) at the beat between the two carriers, $\omega_a=\omega_2-\omega_1$, where $\omega_a$ is the acoustic frequency of the traveling AOM wave. The pulse train acts as a sampling function at the laser repetition frequency $\omega_s$, to generate the beat between sampling frequency and carrier beat $\omega_b\equiv|\omega_a-\omega_s|$ (hence the ``double-beat'' designation). In nonlinear response, because of either nonlinear excitation or nonlinear emission, current appears as harmonics of the beat. The real-time profile of the emission can be reconstructed from the simutaneously measured $n$-harmonics of the cross-correlation current $j_{1,2}^{(n)}(\tau)$, where $\tau$ is the delay between pulse trains.  The scheme maps the optical phase and frequency from PHz to kHz, where current detection with pA sensitivity is possible using current amplifiers.

The nonlinear response of the photocurrent to laser irradiation is summarized by the spectrum in Fig.~\ref{fig:1}(b). Four harmonics of the beat are seen at a nominal irradiation intensity of what may be regarded as a weak field: $I\sim2\times10^{13}$ W/m$^2$. $j_{1,2}^{(n)}(\tau)$ recorded by a lock-in amplier at the four harmonics of the beat, is shown in Fig.~\ref{fig:1}(c). The Fourier transform of the \textit{n}-th order correlation appears at the \textit{n}-th harmonic of the optical carrier, $n\omega_0$: the fourth harmonic of the field, which has a period of 680 attoseconds, appears as electric current modulated at $\sim4$ kHz. The absence of narrowing in the pulse envelopes with harmonic order and their characteristic intensities give away the mechanism of the nonlinear process. The observed set of cross-correlation harmonics can be reproduced assuming coherent tunneling of excited electron populations through the field-modulated barrier:
\begin{multline}
j_{1,2}(\tau;n\omega_0)\propto \int\limits_{-\infty}^\infty [E_1(t)E_2(t+\tau)+c.c.]^n e^\frac{-\phi_0^{(n)}}{|E_1(t)+E_2(t+\tau)|} dt.
\label{current}
\end{multline}
The emission intensities of the harmonics seen as the peaks in the Fourier transforms in Fig. 1c, perfectly fit the Bessel series for $J_n(0.6)$ for $n$=1-4. The harmonic-specific fitting parameter, $\phi_0^{(n)}$, which represents the effective work function for photoexcited electrons, is confirmed by the laser power dependence and the angular distributions in cross-polarized excitation (see below). Note, nonlinearity also arises from the time structure in the exponentiated field correlation (Eq. \ref{current}), which implies that the emission follows the optical phase of the applied field. Evidently, the tunneling time $\tau_T$ is short in comparison to the optical cycle, $\omega_0\tau_T\sim1$, which is the Keldysh adiabaticity criterion that defines the strong field-limit in photoemission \cite{Keldysh}. 

\begin{figure}[t]
\centering
\includegraphics[width=1\columnwidth]{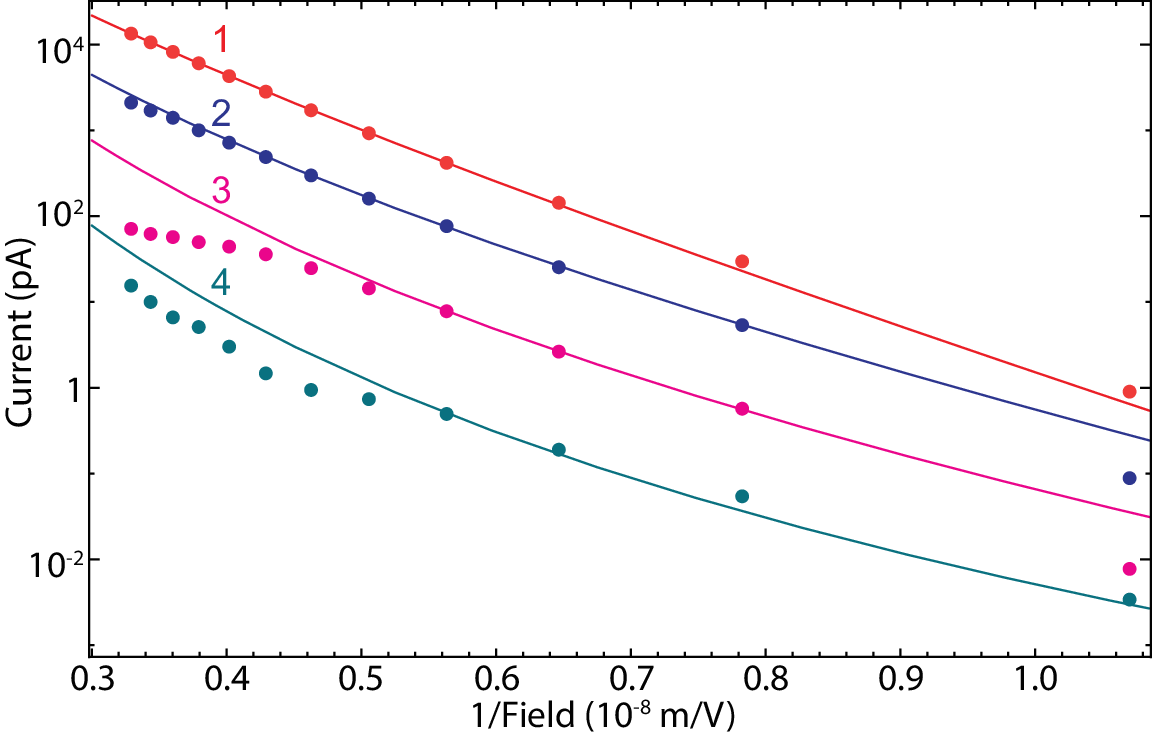}
\caption{\label{fig:2} Semilog plot of the four harmonics of the field emission as a function of inverse applied electric field. Each curve is fitted using Eq. \ref{current}}
\end{figure}

The hallmark of strong-field emission is its exponential dependence on the applied field. This is verified in measurements of current with coincident pulse trains in Fig.~\ref{fig:2}. The data span seven decades in current, corresponding to yields of $10^{-4}$--$10^3$ electrons/pulse. The applied laser intensity spans a narrow range, with upper limit determined by the damage threshold of the apex. Because the excited density of states associated with 3rd and 4th harmonics stretch over the barrer and are likely to be dynamically bound depending on the phase of the time-dependent barrier, the data points obviously deviate from the expected trend at high intensity. However, the data points of third and forth order in the low field limit and entire first and second order, the current obeys an multiphoton excitation-exponential tunneling law, $\ln(j_{1,2}^{(n)}/I)\propto I^n -\phi_0/\sqrt{I}$, instead of the pure perturbative multiphoton limit $j^{(n)} \propto I^n$. Nevertheless, the applied field ($\approx 10^8$ V/m) is two orders of magnitude smaller than the required field to barely meet the adiabaticity criterion of $\gamma \approx 1$. In our case, $\gamma=\omega_0\tau_T=\omega_0\sqrt{2m_e\phi}/(eE_0)\sim100$, where $\phi$ and $E_0$ are work function and applied field, respectively \cite{Keldysh}. The apparent contradiction is resolved by recognizing that the local field is enhanced by a factor $\beta=E/E_0$. If we assume a triangular tunneling potential, as is made in exponentially accurate electrostatic Fowler-Nordheim (FN) \cite{Fowler, Gadzuk} and dynamical theories \cite{Delone}, the effective work function constant in (\ref{current}) can be identified as $\phi_0=4\sqrt{2m_e}\phi^{3/2}/3e\hbar\beta$. Accordingly, for the case of the first order response, the net tunneling emission rate can be quantified as:
\begin{equation}
w=j/e=\frac{\epsilon_0 c \sigma E_0^2}{2 \hbar\omega}\exp(-4\sqrt{2m_e}\phi^{3/2}/3e\hbar\beta E_0),
\label{rate}
\end{equation}
where $E_0 \propto \sqrt{I}$ is the applied field. The nearly perfect fit to the first and second order (Fig.~\ref{fig:2}) confirms the conclusion obtained in the time-harmonics analysis and yields: $\sigma=1$ \AA$^2$ for the cross section and $\beta=100$.

\begin{figure}[b]
\centering
\includegraphics[width=1\columnwidth]{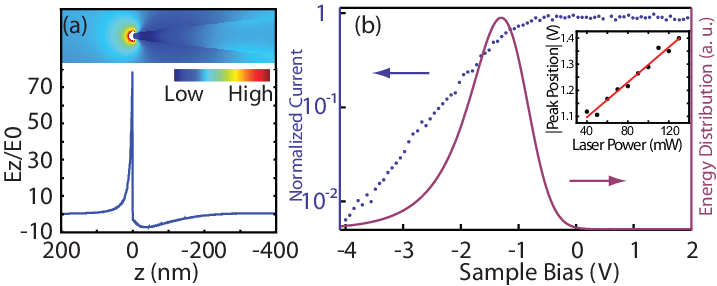}
\caption{\label{fig:3} (a) Computed enhancement of $E_z$ on a silver tip (cone diameter = 20 nm). $\beta=E_z/E_0$ is $\sim$80 at the apex. (b) Bias dependence of the normalized current and the extracted electron energy spectrum at $I=2\times 10^{13}$ W/m$^2$. The inset shows the linear shift of the spectral peak as a function of applied laser intensity, as predicted for ponderomotive acceleration.}
\label{fem}
\end{figure}

For tip shapes in our experiments, $\beta(z=0)\sim100$ is typical. For a tip with 20 nm cone diameter, we find $\beta=E_z/E_0=80$ through finite element analysis as shown in Fig.~\ref{fig:3}(a) \cite{note2}. The addition of a 1 nm asperity yields $\beta=200$. The decay profile of the field into vacuum perfectly fits that of a point-dipole $\beta (z) = \beta (0)/(z+z_0)^3$ for $z>0$, placed at $z_0=15$ nm inside the cone. The extracted cross section suggests that the field acts on the apex mode of the tip plasmon \cite{Novotny2003}. Additional features of the enhancement profile that deserve note are: a) the impressed field is discontinuous. It changes sign at the metal/vacuum interface due to the displaced charge density---the field has a cusp at the vacuum/metal interface. b) Inside the metal, the field decays (skin depth of 20 nm) with a profile that follows the charge displacement along the taper. Clearly only surface electrons, confined to a layer much thinner than the skin depth, can follow the light field in phase to emit coherently. Since electrons excited within the cone are subject to a force $\pi$ out of phase with the field on the vacuum side, they will scatter on the interfacial potential and subsequently thermalize, as previously demonstrated in nano-rods \cite{Zadoyan}. c) At the apex, $\beta$ decays into vacuum on a scale $\l_{1/2}\sim4$ nm (much shorter than the wavelength of light). This steep field gradient ensures ponderomotive acceleration of emitted electrons and imposes multipolar coupling to far field radiation, which we verify respectively through the energy distribution of the collected electrons and their polarization dependence.

\begin{figure}[t]
\centering
\includegraphics[width=1\columnwidth]{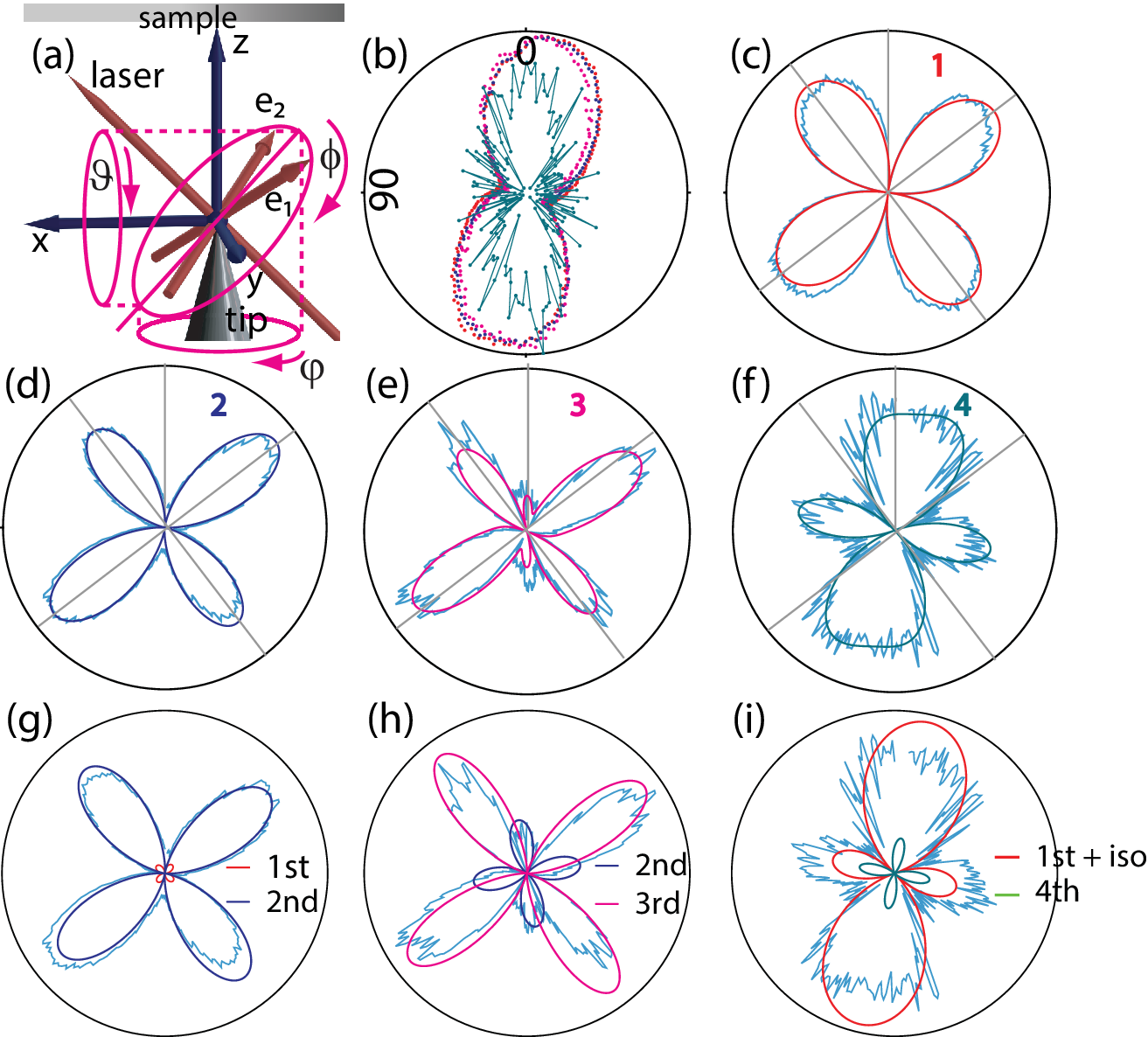}
\caption{\label{fig:4} Angular distribution of photocurrent. (a) Geometry of cross-polarized illumination: the tip is along $\hat{z}$; the propagation is along $\hat{x}$+$\hat{z}$; the polarization angle $\phi$ is zero when $\hat{e_{1}}$ and $\hat{e_{2}}$ are at $\pm 45 ^\circ$. (b) In parallel polarization, the normalized angular distributions of the harmonics are superimposable. (c-f) Angular distributions of 1-4th harmonics in cross-polarized excitation. (g) Decomposition of 2nd harmonic in (d) fitted with $a_1:a_2 =1:23$. (h) Decomposition of 3rd harmonic in (e) with $a_2:a_3 = 1:5.4$. (i) Decompostion of 4th harmonic in (f) with $a_1:a_4:b_4 = 1:-2.6:0.21$. In both polarization configurations, the tip is slightly tilted with respect to the laboratory $z$-axis.}
\end{figure}

The kinetic energy distribution of the detected electrons is characterized through the bias dependence of the photoemission current, $j(V_b)$. The data shown in Fig.~\ref{fig:3}(b) is characteristic for all harmonics. Note, at the junction gap of 1--100 $\mu$m, the static field due to the bias is several orders of magnitude smaller than the light field. It plays no role in the extraction of electrons, but can sweep emitted electrons and provides a retarding potential for energy analysis. The observed current decreases exponentially for $V_b<-1$ V and remains constant above $V_b\sim0$ V. The electron energy spectrum obtained by numerically differentiating $j(V_b)$ peaks at $\varepsilon_z=-eV_b\sim1.3$ eV and decays sharply by $V_b=0$. This peak shift corresponds to the ponderomotive potential of the local field, $U_p=e^2 I/4 m_e \omega_0^2=e^2 \beta^2I_0/4 m_e \omega_0^2$ for $I_0=2\times10^{13}$ W/m$^2$ and $\beta^2=10^4$, $U_p=1.2$ V and shifts linearly with the applied intensity (Fig.~\ref{fig:3} inset). Despite the short duration of the accelerating field, this energy is attained due to the short range ($\sim$4 nm) of the local field, as verified through explicit trajectory calculations (supplement) \cite{alejandro}, following Corkum's two-step semi-classical treatment \cite{Corkum}. Despite the photoexciation of electrons in the tip, structureless, exponentially decaying spectra in all harmonics are observed. If a strong static field is employed, the prepared electron distribution can be preserved \cite{Ropers2007, Yanagisawa2010}. Our energy distributions are smeared by pondermotive acceleration, which obscures the prepared distributions of photoexcited electrons.\\
The dependence of the cross-correlation current and its harmonics on the polarization of the light provides independent verification of the governing mechanism. The experimental irradiation geometry is presented in Fig.~\ref{fig:4}, along with two sets of angular patterns obtained as a function of the angle $\phi$ of the incident electric fields. In the first set, the polarization of the two pulse trains is parallel; in the second set, they are orthogonal (cross-polarized). In the parallel set, the normalized angular distributions of the first three harmonics are identical, while the fourth harmonic pattern is characteristically jagged. These distributions fit to the sum of a dipolar and isotropic term, $6\cos^2(\phi)+1$. They can be understood by considering the transition matrix elements of the linear interference induced on a point-polarizable density, with nonlocality limited to first order in spatial dispersion:
\begin{equation}
\begin{split}
&\langle P_z^{(2)}\rangle\propto\langle n|\pmb E_2(r)\cdot\nabla|m\rangle\langle m|\pmb E_1(r) \cdot\nabla|n\rangle\\
&\propto E_0^2\langle n|(1+i\hm{k\cdot r})(\hat{\varepsilon}_2\cdot\nabla)|m\rangle\langle m|(1+i\hm{k\cdot r})(\hat{\varepsilon}_1\cdot\nabla)|n\rangle.
\label{polarization}
\end{split}
\end{equation}
Dipolar excitation arises from the optical field, induced by the $\hat{x}$-component of propagation where $\phi$ reduces to the projected polar angle $\vartheta$ onto the $yz$ plane. The two fields mix along the $z$-axis giving rise to the second order polarization $P^{(2)}_{z}$ as $\langle d_zd_z\rangle=\langle(\hat{\varepsilon}\cdot\hat{z})(\hat{\varepsilon}\cdot\hat{z})\rangle \propto \cos^2(\vartheta)$. The isotropic term arises from the component along the tip $\hat{z}$-axis, where $\phi$ reduces to the azimuthal angle, $\varphi$, under which $P_z$ is invariant. This represents the quadrupolar excitation $\langle q_{zx}q_{zx}\rangle$ induced by the field gradient $(\partial E/\partial z) / E_0=k_z$ \cite{note4}, characteristic of tip plasmons \cite{Novotny2002} as shown in Fig. \ref{fem}. The absence of narrowing in the harmonic-resolved angular distribution of the parallel set is the outcome of competition between the narrowing nonlinear excitation and broadening exponential tunneling probabilty as the work function is reduced as in the case of cross-correlation traces (See supplement).

The multiphoton absorption is clarified in the cross-polarized geometry. The observation of progressively narrowing lobes in this dataset as the harmonic order increases means that the preexponent, $E_0^{2n}$ dominates while the exponential tunneling probabilities for all harmonics are nearly saturated due to the strong field enhancement. They arise undoubtedly from photoexcited populations with distinct effective workfunctions, and the fits are performed with the same time-dependent FN equation as such, with the field changing in the polarization angle. The dipolar excitation appears as a clover pattern, $|\cos(\vartheta)\cos(\vartheta+\pi/2)|=|\sin(2\vartheta)|/2$, and fits well the 1st harmonic distribution which is consistent with our prior report in which the CPDB method is implemeted in the STM imaging of a single Ag atom \cite{Lee}. The 2nd and 3rd harmonics show uneven clovers with narrower lobes, indicative of $\sin^2(2\vartheta)$ and $\sin^3(2\vartheta)$ contributions, respectively. The broken symmetry is explained by interference between different orders. The decomposition of angular dependence reveals that there is crosstalk between different orders.
The basis functions for fitting each harmonic angular distribution are:
\begin{equation}
\begin{split}
& A \left|\sum_n a_n \cos(\theta +\alpha_n)^n\cos(\theta+\pi/2 +\alpha_n)^n+b_n\right|,\\
\end{split}
\end{equation}
where $A$ is the amplitude, $n$ is the harmonic order, $a_n$ is the weight factor of $n$th order absorbance, $b_n$ is the isotropic term for fitting $n$th order current, $\theta$ is the polarization angle, and $\alpha_n$ is the angular offset from nominal polarization angle for each harmonic. 
The product function accounts for the coherence of two cross-polarized pulses and their individual absorbance into the tip.  
The slightly uneven lobes of the 2nd harmonic are explained by two constructively interfering and two destructively interfering lobes between 1st and 2nd harmonic components. The first order-prepared population obviously appear in the second harmonic response due to the nonlinearty of tunneling. The 3rd order angular distribution is decomposed into the 2nd and 3rd order populations, with 2nd order preparation is contributing to the 3rd order emission through nonlinear tunneling. The 4th harmonic pattern is clearly reduced to two-fold symmetry and reproduced by the interference between 1st, 4th, and an isotropic emission. It is noteworthy that the 4th harmonic emission is orthogonal to other three harmonics. The polarization mixing is maximized when both arms have maximum projection (at $\theta=\pm \frac{\pi}{4}$) along the tip axis in which the anisotropic susceptibility of plasmon is maximized. The orthogonal emission means the emission takes place when one arm polarization is along the tip and the other is perpedicular, which leads to nearly zero mixing for other three harmonics. However, there is a small istropic component of susceptibility at the tip apex. The cross-polarization minimally mixes at this angle but the 4th harmonic takes advantage of nearly zero workfucntion, emitting in both normal and parallel directions with respect to the tip surface as shown in Fig.~\ref{fig:4}. Additionally, since the energy of electrons modulated at the 4th harmonic of the optical carrier is above the vacuum level, they only exist in dynamically field-bound states that spill into vacuum, presumably beyond the stationary field states of silver for which $\langle z\rangle>1$ nm  \cite{Echenique, Kliewer}. 
\begin{comment}
The high harmonics will likely be generated in high local field, which is consistent with our observation showing multipolar fields in 2-4th harmonics. It is possible to estimate the amplitude of the induced displacements. From the relative magnitude of dipole and quadrupole in parallel excitation, $\langle k_zz\rangle=2^{-1/2}$, using the computed range parameter of the local field,  $1/k_z=\ l_{1/2} \sim 4 $ nm, we may estimate $\langle z\rangle \sim 3$ nm.
\end{comment} 
The different orders in nonlinear response can be associated with increasingly larger amplitude displacements.

A rather complete description of coherent field emission emerges from the measurement set. The process can be uniquely ascribed to tunneling detachment of the plasmon wavefunction that spills out into vacuum, on the steep gradient of the self-enhanced local field. This collective mechanism should hold quite generally for plasmonic nano-structures and suggests novel applications such as ultrafast imaging of adsorbates on surfaces with subcycle emission in properly designed fields and tip geometries. \\

This research was made possible through the NSF Center for Chemistry at the Space-Time Limit (Grant no. CHE-082913). A.R. is grateful for the NSF graduate research fellowship (Grant No. DGE-0808392).

\bibliographystyle{apsrevM}
\bibliography{fieldemission}

\end{document}

\preprint{APS/123-QED}

\title{Supplementary Material:\\ Coherent field emission upon ultrafast laser irradiation of the tip plasmon}% Force line breaks with \\

\author{Joonhee Lee}
\affiliation{Department of Physics, University of Nevada, Reno, Reno, Nevada, 89557, USA}
\author{Shawn M. Perdue}
\affiliation{Department of Chemistry, University of California, Irvine, Irvine, California, 92697-2025, USA}
\author{Alejandro Rodriguez Perez}
\affiliation{Department of Chemistry, University of California, Irvine, Irvine, California, 92697-2025, USA}
\author{V. Ara Apkarian}
\affiliation{Department of Chemistry, University of California, Irvine, Irvine, California, 92697-2025, USA}

\date{\today}% It is always \today, today,

\begin{comment}
\begin{abstract}
An article usually includes an abstract, a concise summary of the work
covered at length in the main body of the article. 
\begin{description}
\item[Usage]
Secondary publications and information retrieval purposes.
\item[Structure]
You may use the \texttt{description} environment to structure your abstract;
use the optional argument of the \verb+\item+ command to give the category of each item. 
\end{description}
\end{abstract}
\end{comment}

\maketitle

%\tableofcontents

\section{\label{sec:level1}Method\protect\\}

\begin{figure}[b]
\includegraphics[width=1\columnwidth]{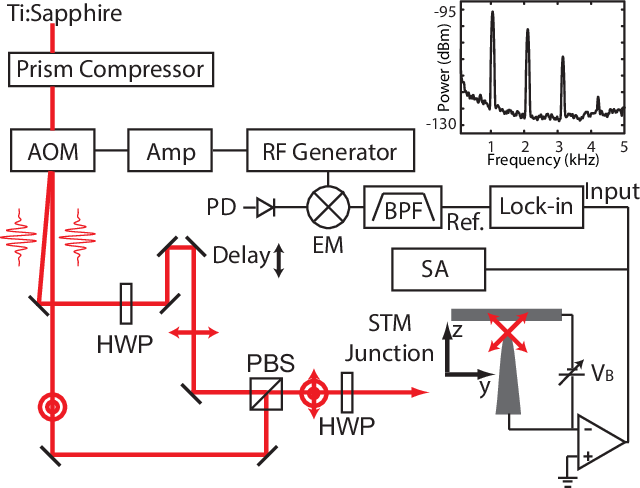}% Here is how to import EPS art
\caption{\label{fig:figs1} Schematic diagram of the experimental setup. AOM: acousto-optic modulator, Amp: RF amplifier, PD: photodiode monitoring the repetition rate of the Ti:Sapphire laser, BPF: electronic band pass filter, EM: electronic mixer, SA: spectrum analyzer, HWP: half-wave plate, PBS: polarizing cube beam splitter. The base-band spectrum of the photocurrent obtained from the spectrum analyzer is shown in the upper right corner. 4 harmonics of the fundamental double beat frequency $f_b$ are clearly resolved.}
\end{figure}

The schematic diagram of the optical setup used in present experiments is shown in Fig.~\ref{fig:figs1}. The output of a Ti$^+$:Sapphire laser is passed through an acousto-optic modulator (AOM), which acts as a frequency shifter. The undiffracted beam and the 1st-order diffracted beam are then cross-polarized, recombined in a Mach-Zehnder interferometer, and passed through a $\lambda/2$-waveplate. The collinear beams are focused on the STM-junction with a plano-convex lens (focal length = 10 cm, nominal spot size = 25 $\mu m$), at 45$^\circ$ incidence relative to the tip axis (z-axis) as shown in Fig. 1(a). The two pulse trains, $\hat{\epsilon}_1 E_1(\omega_1,t)$ and $\hat{\epsilon}_2 E_2(\omega_2,t)$, are distinguished by their mutually orthogonal polarization directions, $\hat{\epsilon}_1 \cdot \hat{\epsilon_2}=0$, and carrier frequencies: $\omega_1 = \omega_o$  and $\omega_2 = \omega_o+\omega_a$, where $\omega_o$ is the optical frequency of the laser and $\omega_a$ is the acoustic frequency of the traveling AOM wave. The electronic coherence prepared in linear excitation is the cross-polarization, which in strictly local approximation is ~[\onlinecite{Svirko}]:.

\begin{equation}
P_i^{(2)}(t)=\chi_{ijk}^{(2)}(\omega_1-\omega_2,\omega_1,-\omega_2)E_{1,j}E_{2,k}^*(t)+c.c.
%P_z^{(1)}(t)=\chi_{zz}^{(1)}[E_{1z}(t)+E_{2z}(t)]
\end{equation}
Two pulse trains in Gaussian envelops overlapping in time and projected onto the tip $z$-axis are expressed as: 
\begin{equation}
\begin{split}
&E_{1z}(t)=\sum_n E_0e^{-(\frac{t-nT}{\Delta t})^2}e^{i(kr-\omega_o t)},\\
&E_{2z}(t)=\sum_n E_0e^{-(\frac{t-nT}{\Delta t})^2}e^{i[kr-(\omega_o+\omega_a) t]},\\
\end{split}
\end{equation}
where $E_0$ is the balanced amplitude of the electric field, $T$ is the repetition period, and $\Delta t \approx$ the FWHM pulse width. %$\hat{\epsilon}_1 = \frac{\hat{z}+\hat{y}}{\sqrt{2}}$ and $\hat{\epsilon}_2 = \frac{\hat{z}-\hat{y}}{\sqrt{2}}$.

Then the $z$-compoment of the cross-polarization is:
\begin{equation}
\begin{split}
&P_z^{(2)}(t) =\chi_{zzz}^{(2)} E_0^2 \sum_n e^{-2 \left( \frac{t - nT}{\Delta t} \right)^2} \cos(\omega_a t).
%&|P_z^{(1)}(t)|^2
\label{p2pol}
\end{split}
\end{equation}
This linear interference consists of a wave modulated at the beat between optical carriers, sampled with $\Delta t\approx$50 fs pulses, at the repetition rate of the laser $f_s = 1/T$ = 76 MHz. The spectrum of Eq. \ref{p2pol} consists of a harmonic comb and sidebands: 
\begin{equation}
\sum_{m,l} \alpha_{m,l} \delta \left[ f - (m f_s \pm f_a) \right],
\end{equation}
where $m=0,1,2, \cdots;$ and $|f_s-f_a| \equiv f_b$ is the beat between sampling frequency and carrier beat (hence the double-beat designation). In $n$-th order response, $P^{2n}$ gives the interference, and the spectral comb develops the additional sidebands \cite{notes1}.
\begin{equation}
\sum_{m,l}^n \alpha_{m,l} \delta \left[ f - (m f_s \pm l f_a) \right],
\label{eq:additionalsidebands}
\end{equation}
where $m=0,1,2, \cdots; l=0,1,2, \cdots, n$.
Various parametric processes map in base-band, defined by $m=l$. For example, in second order response, polarization driven at the sum frequency 2$\omega_0$ (0.75 PHz) appears at the 2nd harmonic of the beat, $2f_b = |2f_s-2f_a|$. For a laser repetition rate of $f_s$ = 76 MHz and AOM driven at an RF frequency of 76 MHz + 1 kHz, the second harmonic response at 0.75 PHz is shifted down to 2$f_b$ = 2 kHz.This double-beat frequency can be arbitrarily controlled by the RF generator, which is actively locked to the laser repetition rate to compensate for cavity length variations. The beat frequency is selected to lie in the kHz range, where current detection with pA sensitivity is possible using amplifiers of typical gain-bandwidth. To the extent that nonlinearties are parametric, phase sensitive detection by locking on harmonics of the beat frequency allows reconstruction of the time profile of the coherence. Or, at the expense of losing phase information, the spectrum Eq.~\ref{eq:additionalsidebands} can be measured directly in frequency domain using a spectrum analyzer. These relate to measuring the polarization with a linear response detector. We detect field emission from the prepared polarization, with strong nonlinearity inherent in tunneling ionization. By expanding the detection response in a power series in the time-dependent local field, the detectable cross-correlated photoemission current can be described with generality: 
\begin{equation}
j_{1,2}(t)
= \sum_{n} \chi^{(2n)} E_1^{\,n}(t)\, E_2^{\,n^\ast}(t)
  \sum_{m} \xi_m^{\,n} \bigl(E_1(t) + E_2(t)\bigr)^{m}.
\label{eq:parametricemission}
\end{equation}
to recognize that parametric mixing can occur due to nonlinearity in excitation, in detection, or both. While $\chi^{(n)}$ susceptibilities are a decreasing power series in $n$, tunneling emission probabilities from $n$-photon excited states, $\xi^{(n)}$, would be expected to increase exponentially with $n$. Independent of the origin of the nonlinearity, experimentally, the time profile of the parametric emission (Eq.~\ref{eq:parametricemission}) is constructed by measuring the harmonics of the cross-correlation current, $j_{1,2}^{(n)}(\tau)$, constructing their phased transforms, then back-transforming the summed spectra,
\begin{equation}
j_{1,2}(t)
= \int\sum_{n} j_{1,2}^{(n)}(\omega)e^{i\omega t}d\omega.
\label{eq:xcorrcurrent}
\end{equation}
The controlling nonlinearities are sorted out through auxiliary information: intensity dependence, angular distribution and energy spectrum of the emitted electrons. In the main text, we show that Eq.~\ref{eq:parametricemission} with $n$ = $1, 2, 3, 4$ and the exponential series in $m$ explains the nonlinear excitation and nonlinear emission controlled by coherent tunneling, in the case of cross-correlation current and power dependence.

\section{\label{sec:level1}Simulated cross-correlation of each harmonic current}
The cross-correlation traces are calculated by Eq. \ref{currentsim}. 
\begin{equation}
j_{1,2}(\tau)\propto (E_0 E_0^*)^{n}\int\limits_{-\infty}^\infty e^{-\frac{\phi_0^{(n)}}{\sqrt{|E_0 E_0^*|}}}dt .
\label{currentsim}
\end{equation}
where the sum of two optical fields, $E_0 =E_1(t)+E_2(t+\tau)$ and $n$ is the harmonic order. The effective workfunctions extracted from the power dependence data are: $\phi^{(1)}$ = 3.19 eV, $\phi^{(2)}$ = 2.22 eV, $\phi^{(3)}$ = 0.37  eV, and $\phi^{(4)} = 6.5 \times 10^{-8}$ eV. The $n$th order excitation in the preexponent narrows whereas the reduced workfunction in higher order broadens the electron pulses. In effect, the simulated traces do not exhibit significant narrowing as the harmonic order increases as shown in the experimental dataset in the main text.  In the case of optical autocorrelation, the trace narrows progressively as the nonlinear order increases.

\begin{figure}[h]
\includegraphics[width=0.8\columnwidth]{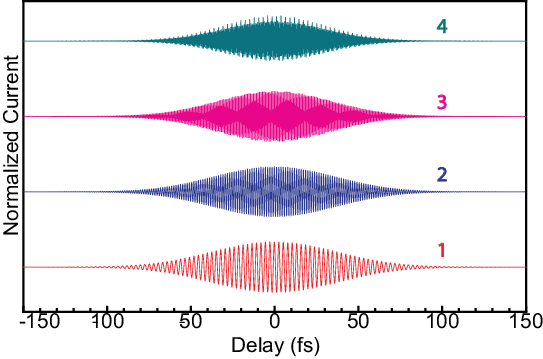}% Here is how to import EPS art
\caption{\label{fig:Simulated cross-correlation traces.} Simulated cross-correlation traces. The incident laser power is 41 mW and the enhancement factor is $\approx$80.}
\end{figure}

\section{\label{sec:level1}Electron Energy Distribution}
\begin{figure}[b]
\includegraphics[width=0.75\columnwidth]{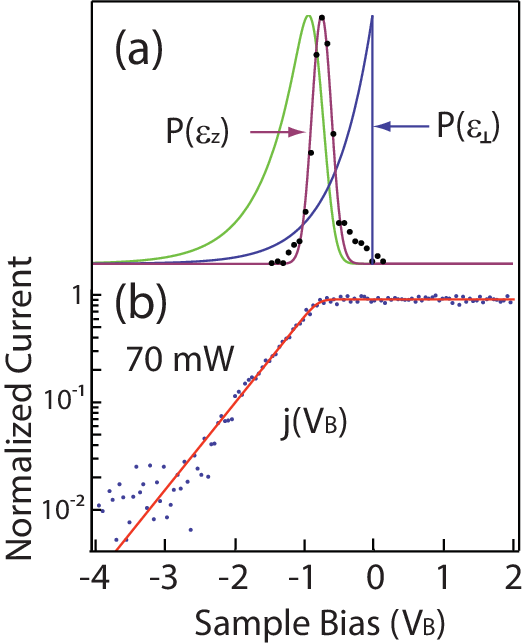}% Here is how to import EPS art
\caption{\label{fig:energydist} Kinetic energy distribution of emitted electrons as a function of sample bias $V_B$. (a) Normal energy distribution curve $P(E_{\perp})$, in blue, and the computed distribution of electron energies accelerated by the field $P(E_z)$, in purple. Convolution of $P(E_{\perp})$ and $P(E_z)$ yields the differential electron energy distribution (green). (b) Integrated differential current is used to compare with the 1st harmonic experimental data at 70 mW laser irradiation.}
\end{figure}

The observed electron energy distributions can be quantified by following Corkum’s two-step semi-classical model for tunneling ionization \cite{Corkum}. We assume that an electron is born in vacuum under the laser pulse, at $t = t_i$, with probability $w(t_i)$ given by Eq. 2 in the main text, and including the explicit time dependence of the field. We compute their final energy by propagating trajectories subject to the time dependent local field, $F(t)=\beta(z)E(t)\cos(\omega_o t)$, using the numerically obtained field profile $\beta(z)$ and under the Gaussian pulse envelope $E(t)$ of the laser. The obtained distribution, $P(E_z)$, is then convoluted with the exponential normal energy distribution $P(E_{\perp})=\exp(-E_{\perp}/d)$ of standard FN theory \cite{Young}, to generate the total energy distribution. The experimental distribution can always be fit by an exponential-Gaussian convolution (Fig.~\ref{fig:energydist}). Notwithstanding the approximations, we may conclude that electrons accelerate post ionization by the gradient of the field-potential of the enhanced local field of the plasmon. 

\section{\label{sec:level1}Electric quadrupole and magnetic dipole}
The interaction hamiltonian ~[\onlinecite{Schatz}],
\begin{equation}
\begin{split}
&V_{mn} = -i\frac{eA_0}{2mc}\bra{m}(\pmb k \cdot \pmb r)(\pmb\epsilon \cdot \pmb p)\ket{n}\\
&=-\frac{iA_0}{2}(\pmb k \times \pmb \epsilon)\cdot\pmb M_{mn}+\frac{A_0\omega_{mn}}{4ce}\pmb \epsilon \cdot Q_{mn} \cdot \pmb k,\\
&\text{where } \pmb M = \frac{e}{2mc}\pmb L, \text{ and } (Q_{mn})_{ij}=\bra{m}(er_i)(er_j)\ket{n}, 
\label{eq:inthamt}
\end{split}
\end{equation}
where $\pmb k$ is the wavevector of light, $\pmb \epsilon$ is the polarization vector, and $\pmb M$ is the magnetic dipole moment. Consider the projection of light propagation onto the z-axis. For $\pmb k = k_0 \hat{k}$ and $\pmb \epsilon =\hat{i}$, Eq.~\ref{eq:inthamt}  becomes the following:
\begin{equation}
\begin{split}
&-i\frac{eA_0}{2mc}\bra{m}(k_0 z)(p_x)\ket{n} \\
&=-\frac{ieA_0}{4mc}(k_0\hat{j})\cdot\bra{m}\pmb L\ket{n}+\frac{eA_0\omega_{mn}k_0}{4c}\bra{m}xz\ket{n}.\\
\end{split}
\end{equation}
Therefore,
\begin{equation}
\bra{m}\hat{z}\hat{\partial_x}\ket{n} = \frac{i}{2\hbar}\bra{m}L_y\ket{n}-\frac{m\omega_{mn}}{2\hbar}\bra{m}xz\ket{n}.
\end{equation}

\begin{comment}
\section{Polarization Data Analysis}
In the case of polarization data, the current modulation as a function of polarization angle is moderate compared to the case of interferometric scan. Therefore, the angular dependence is dominated by the preexponent, representing the nonlinearly prepared excited population. 

\begin{figure}[h]
\includegraphics[width=0.8\linewidth]{polarization_breakdown.eps}% Here is how to import EPS art
\caption{\label{fig:energydist} Fitting parameters for polarization dependence of photocurrent. (Top): Second harmonic fitted with $a_1:a_2 =1:23$.  (Middle) Third harmonic. $a_2:a_3 = 1:5.4$. (Bottom) Fourth harmonic: $a_1:a_4:b_4 = 1:-2.6:0.21.$}
\end{figure}

%\begin{widetext}
%\end{widetext}

The basis functions for fitting each harmonic current are:
\begin{equation}
\begin{split}
& A \left|\sum_n a_n \cos(\theta +\alpha_n)^n\cos(\theta+\pi/2 +\alpha_n)^n+b_n\right|,\\
\end{split}
\end{equation}
where $A$ is the amplitude, $n$ is the harmonic order, $a_n$ is the weight factor of $n$th order absorbance, $b_n$ is the isotropic term for fitting $n$th order current, $\theta$ is the polarization angle, and $\alpha_n$ is the angular offset from nominal polarization angle for each harmonic. 
The product function accounts for the coherence of two cross-polarized pulses and their individual absorbance into the tip.  
\end{comment}

\bibliography{supplement}% Produces the bibliography via BibTeX.